# Closing Africa's Early Warning Gap: AI Weather Forecasting for Disaster Prevention

**Qness Ndlovu**

*Dimension Research Lab, Cape Town, South Africa*

research@dimensionresearchlab.com

## Abstract

In January 2026, torrential rains killed 200–300 people across Southern Africa and affected over 700,000, with South Africa declaring a national disaster on January 17. Despite meteorological agencies issuing warnings, the crisis exposed a critical reality: 60% of Africa's population lacks access to effective early warning systems [1], contributing to disaster death rates 4x higher than Europe and North America [2]. The structural barrier is cost – traditional radar infrastructure exceeds $1 million per station, limiting the continent to just 37 radar installations for 1.2 billion people, compared to 636 stations serving 1.1 billion in the US and EU combined.

We present a production-grade architecture for deploying NVIDIA Earth-2 AI weather models at $1,430–$1,730/month for national-scale deployment – enabling early warning coverage at 2,000–4,545× lower cost than traditional radar infrastructure – while generating 15-day global atmospheric condition forecasts; periodic GPU forecasts are cached in PostgreSQL, enabling user-facing queries under 200 milliseconds without real-time inference. Our system, deployed in February 2026 to evaluate how Earth-2-based forecasting could be operationalized, demonstrates three technical contributions: (1) a ProcessPoolExecutor-based event loop isolation pattern that resolves aiobotocore session lifecycle conflicts in async Python applications, enabling reliable GFS data ingestion; (2) a database-backed forecast serving architecture where the GPU writes global forecasts directly to PostgreSQL and CPU servers query location-specific subsets in under 100 milliseconds, eliminating HTTP transfer bottlenecks for high-resolution atmospheric tensors; and (3) automated coordinate management pattern for multi-step inference across 61 timesteps, eliminating manual dimension tracking failures in production deployments. The system delivers forecasts through WhatsApp-based distribution architecture, reaching users through an 80%+ penetration communication channel. This architecture was deployed at national scale in South Africa.

Countries with effective early warning systems experience disaster death rates 6x lower than those without. By reducing deployment costs from hundreds of millions to tens of thousands of dollars annually – achieving national-scale coverage at 2,000–4,545× lower cost than traditional radar infrastructure – this architecture enables continent-scale replication of early warning systems. UNDRR independently documents that such systems reduce disaster death rates 6x [4]. All architectural details and implementation patterns are documented inline to ensure full reproducibility.

## 1. Introduction

### 1.1 The Human Cost

Between 1970 and 2019, weather-related disasters in Africa caused 731,747 deaths across 1,695 events [1]. The crisis is accelerating: Africa experienced its deadliest five-year period (2021–2025) in over 15 years, with extreme weather disasters claiming nearly 29,000 lives and affecting over 221 million people across the continent [7]. In 2023 alone, extreme weather events killed 15,700 people and impacted 34 million across the continent [8].

The January 2026 Southern Africa flooding event exemplifies the ongoing toll. Beginning in mid-December 2025, severe rainfall systems affected six countries across the region: Mozambique, South Africa, Zimbabwe, Malawi, Eswatini, and Madagascar. By January 17, 2026, South Africa declared a national disaster after floods killed at least 37 people, destroyed over 1,000 homes, and caused extensive damage to roads, bridges, and public infrastructure across five provinces – Limpopo, Mpumalanga, KwaZulu-Natal, Eastern Cape, and North West – with Limpopo and Mpumalanga bearing the brunt of the devastation [9, 10, 11]. The South African Weather Service issued a rare Red Level 10 warning – their highest alert category – yet evacuations of the Kruger National Park and surrounding communities came too late for many [12, 13]. Across Mozambique, over 600,000 were affected with more than 50,000 displaced. Zimbabwe reported 70 fatalities. The regional death toll reached 200–300 people, with over 700,000 affected across Southern Africa [14, 11, 15].

Despite the South African Weather Service issuing accurate Red Level 10 forecasts, evacuations of Kruger National Park occurred reactively after flooding began, and hundreds of guests required emergency extraction [12, 13], highlighting gaps in translating forecasts into timely community action [1, 7, 9]. The technical capacity to predict existed; the systemic capacity to act did not.

> *Note: This paper was prepared in February 2026 using the January flooding event as motivating context. Post-event analysis cited in this section draws from preliminary reports and may be updated as official assessments are completed.*

## 1.2 The Infrastructure Barrier

The disparity in early warning infrastructure is stark. The United States and European Union operate 636 weather radar stations serving 1.1 billion people – approximately one radar per 1.7 million residents [16]. Africa operates 37 radar stations for 1.2 billion people – one radar per 32 million residents, an 18x coverage deficit (Figure 1) [16]. As of 2019, only 20% of African weather stations met World Meteorological Organization standards [16].

The barrier is economic, not technological. A single state-of-the-art weather radar installation costs $2–8 million [17, 18], reflecting differences in technology sophistication (basic vs. dual-polarization Doppler), infrastructure requirements (towers, power, communications), and installation complexity. South Africa's Weather Service invested R240 million (~$33M USD at 2010 exchange rates) to deploy 12 Doppler radars [17], averaging $2.75 million per installation, while international procurement of advanced WSR-88D systems has exceeded $4.77 million per unit [18]. Multi-radar networks require $25–250 million in capital expenditure plus $1 million annually in maintenance. The World Bank estimates $1.5 billion is needed for continent-wide hydromet systems in Africa, which would save African countries $13 billion in asset losses and $22 billion in livelihood losses annually [19]. Just 24 hours' notice of an impending hazardous event can reduce damage by 30 percent [19]. Meanwhile, of every US$100 in official development assistance between 2010–2019, only 50 cents was invested in preventive disaster risk reduction [20].

Yet the return on investment is unambiguous. Countries with effective early warning systems experience disaster death rates 6 times lower than countries without such systems [2]. These benefits must be weighed against climate change costs projected to exceed $50 billion annually across Africa by 2050 [21]

**Figure 1**

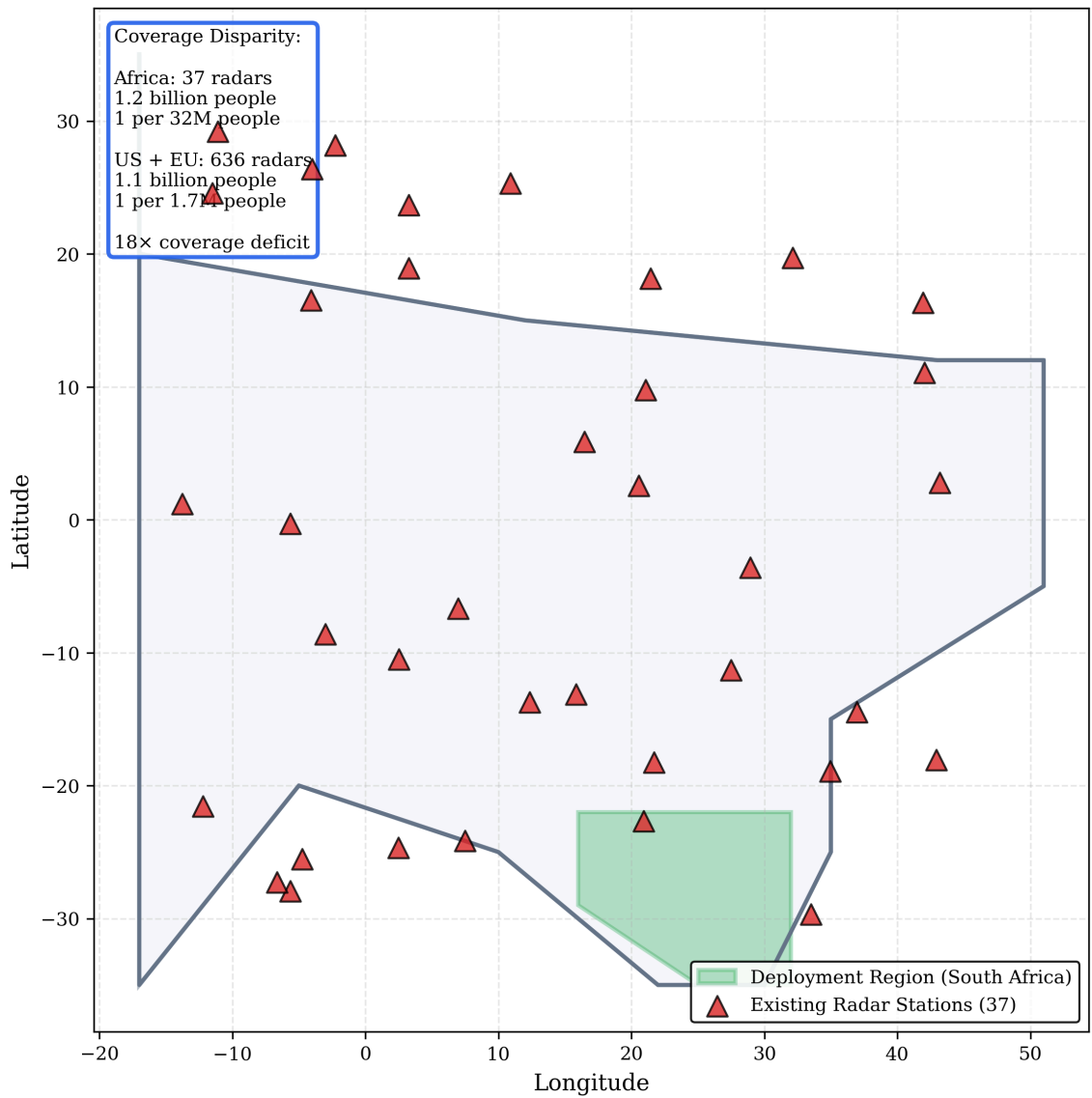

*Figure 1:* Geographic coverage disparity across Africa. With only 37 radar stations serving 1.2 billion people (1 per 32 million), Africa faces an 18× coverage deficit compared to the US and EU combined.

The deaths in Limpopo, Gaza, and across the region in January 2026 were not inevitable. They were the product of an infrastructure gap that economics have made politically unsolvable – until now.

## 1.3 The AI Weather Revolution

In January 2026, NVIDIA released Earth-2 Studio [3], a production-grade platform for AI-based weather forecasting. Unlike traditional numerical weather prediction, which solves atmospheric physics equations on supercomputers, AI models learn atmospheric dynamics directly from decades of reanalysis data. Models like GraphCast [4], FourCastNet [5], Atlas (NVIDIA, 2026), and DLWP [6] achieve competitive accuracy at a fraction of computational cost – seconds of GPU inference versus hours of supercomputer time.

However, existing literature focuses almost exclusively on model accuracy benchmarks. Questions of real-world deployment – infrastructure cost, data movement, reliability, last-mile delivery – remain largely unaddressed. This paper fills that gap.

## 1.4 Research Question

*Can AI-based weather forecasting be operationalized in resource-constrained settings at costs orders of magnitude lower than traditional infrastructure, while maintaining sufficient accuracy and reliability for disaster early warning?*

We evaluate this question through a deployment architecture implemented in South Africa, demonstrating the feasibility of generating 15-day global atmospheric condition forecasts at $1,430–$1,730/month for national-scale deployment – achieving 2,000–4,545× lower cost than traditional radar infrastructure. By reducing deployment costs from hundreds of millions to tens of thousands of dollars annually, this architecture enables continent-scale replication of early warning systems in the 60% of Africa currently lacking any early warning infrastructure. UNDRR independently documents that such systems reduce disaster death rates 6x [2]. This paper documents the technical architecture, deployment patterns, and engineering decisions enabling this cost reduction.

# 2. Related Work

## 2.1 Traditional Meteorological Infrastructure

Conventional early warning systems rely on dense networks of surface weather stations, upper-air observations, radar installations, and satellite data ingested into numerical weather prediction models. While effective in high-income regions, capital requirements and maintenance complexity create structural barriers in Africa. Radar coverage gaps leave 60% of the continent's population without reliable warnings [1].

## 2.2 AI-Based Weather Foundation Models

Recent advances in deep learning have produced weather models capable of learning atmospheric dynamics from reanalysis datasets. GraphCast [4] (DeepMind, 2023), Pangu-Weather [23] (Huawei, 2023), FourCastNet [5] (NVIDIA, 2022), Atlas (NVIDIA, 2026), and DLWP [6] demonstrate competitive or superior performance to traditional numerical approaches, with inference times measured in seconds rather than hours. Independent benchmarks validate the accuracy foundation this architecture inherits: GraphCast outperforms ECMWF HRES – the operational gold standard – on 99.7% of tropospheric test variables [4]; Google DeepMind's subsequent GenCast model outperforms ECMWF ENS, the ensemble system used by 35 nations, on 97.2% of forecast targets at lead times exceeding 36 hours [22]; and NVIDIA's Atlas outperforms GenCast on key industry benchmarks [3]. This architecture does not merely claim sufficient accuracy for early warning – it builds on a model lineage that has surpassed the systems national meteorological agencies currently rely upon. NVIDIA's Earth-2 Studio [3] (2026) provides a production framework for deploying these models, but does not prescribe deployment architectures for constrained environments.

## 2.3 Last-Mile Communication in Disaster Contexts

Even where forecasts exist, dissemination remains a bottleneck. Smartphone applications and web dashboards fail to reach significant portions of African populations. WhatsApp, however, exhibits penetration rates of 80–95% among South African smartphone users (over 28 million monthly active users in a population of 60 million [24]), with similarly high rates in countries like Nigeria and Kenya, making it a dominant, familiar platform continent-wide. The January 2026 floods exposed a distribution challenge: despite accurate Level 10 forecasting, failures in last-mile communication – not prediction – amplified devastation [12, 9].

# 3. System Design

## 3.1 Design Goals

We architect the system around four requirements:

- **Affordability:** Operate at 2,000–4,545× lower cost than traditional radar infrastructure at national scale
- **Reliability:** Function under concurrent load and unreliable networks
- **Reproducibility:** Enable independent deployment without proprietary dependencies
- **Accessibility:** Deliver actionable forecasts through high-penetration, low-friction channels

## 3.2 Architecture Overview

We implement a two-tier architecture (Figure 2) separating I/O-bound data operations from compute-bound GPU inference.

**Control Plane (I/O-Bound):**

- Asynchronous ingestion of Global Forecast System (GFS) data from NOAA AWS S3
- Tensor preprocessing and assembly
- User request orchestration via PostgreSQL forecast queries (<100ms per query)
- WhatsApp Business API integration

*Implemented: FastAPI with AsyncIO on Azure VM (5–10 B2s instances at national scale)*

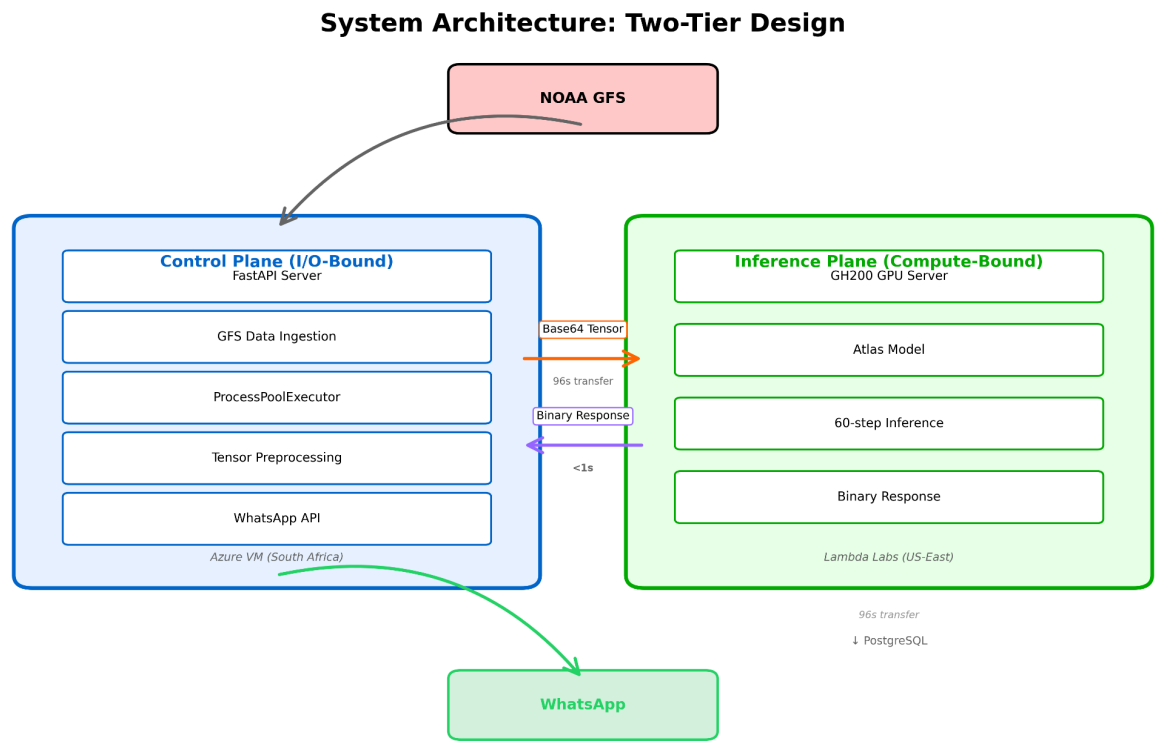

**Figure 2:** *System Architecture showing the two-tier database-backed design. The GPU inference plane generates global forecasts and writes directly to PostgreSQL. The CPU control plane serves user requests by querying location-specific subsets from the database in under 100 milliseconds. Scaling to additional countries requires only additional CPU servers querying the same database.*

**Inference Plane (Compute-Bound):**

- Synchronous Atlas model inference on GPU
- Serialized execution to maximize utilization
- Direct PostgreSQL forecast storage (~1,038,240 rows per global forecast)
- Minimal JSON response protocol (~200 bytes)

*Implemented: Lambda Labs GH200*

This separation enables independent scaling: the CPU-bound control plane handles thousands of concurrent requests by querying forecast data from PostgreSQL, while the GPU plane performs deterministic inference workloads and writes global forecasts directly to the database.

# 4. Technical Contributions

## 4.1 ProcessPoolExecutor Pattern for aiobotocore Event Loop Isolation

**Problem:** Earth-2 Studio's GFS data source uses aiobotocore to fetch atmospheric data from AWS S3. The library manages AWS session lifecycle via async context managers (__aenter__/__aexit__). When GFS fetching is called from within FastAPI's existing event loop, Earth-2's internal use of loop.run_until_complete() creates nested event loop execution. This causes the aiobotocore session to open but never properly "enter" as an async context manager, resulting in cleanup failures with AssertionError: Session was never entered.

**Failed Approach:** Creating a new event loop via asyncio.new_event_loop() within a thread does not isolate the aiobotocore session – both loops share the same process memory space and the session context remains corrupted.

**Solution:** We isolate GFS data fetching in a completely separate Python process using ProcessPoolExecutor. The subprocess has independent memory space, independent event loop, and independent aiobotocore session lifecycle. Data is serialized via pickle, fetched, and returned as a dictionary containing NumPy arrays and metadata.

**Implementation:**

```
from concurrent.futures import ProcessPoolExecutor

def _fetch_gfs_in_process(cycle_time_iso: str, variables: list,
        source: str = "aws", cache_dir: str = "./cache",
        async_timeout: int = 900) -> dict:
    """Standalone GFS fetch - runs in separate process via ProcessPoolExecutor
    Fetches all 75 atmospheric variables required by Atlas model"""
    import os
    from datetime import datetime
    from earth2studio.data import GFS
    os.environ['EARTH2STUDIO_CACHE'] = cache_dir
    cycle_time = datetime.fromisoformat(cycle_time_iso)
    gfs = GFS(source=source, cache=True, verbose=True,
async_timeout=async_timeout)
    data = gfs(cycle_time, variables)
    return {
        "values": data.values,
        "shape": list(data.shape),
        "dims": list(data.dims),
        "coords": {
            "time": [str(t) for t in data.coords["time"].values],
            "variable": list(data.coords["variable"].values),
            "lat": data.coords["lat"].values.tolist(),
            "lon": data.coords["lon"].values.tolist(),
        }
    }

# In async context:
loop = asyncio.get_event_loop()
with ProcessPoolExecutor(max_workers=1) as executor:
    result = await loop.run_in_executor(
        executor,
        _fetch_gfs_in_process,
        cycle_time.isoformat(),
        variables,
        "aws",
        CACHE_DIR,
        GFS_ASYNC_TIMEOUT
    )
```

**Performance:** GFS fetch completes in 3 seconds for 75 atmospheric variables (including surface variables like u10m, v10m, t2m, sp, msl, tcwv; pressure level variables for u, v, z, t, q at 13 levels from 50mb to 1000mb; plus sst and tp) covering the entire globe at 0.25° resolution.

**Generalizability:** This pattern applies to any scenario where async AWS SDK operations must be called from within an existing event loop – a common problem in data science pipelines using FastAPI, Jupyter, or other async frameworks.

## 4.2 Database-Backed Forecast Serving Architecture

**Problem:** The full 61-timestep global atmospheric forecast contains approximately 4,749,948,000 floating-point values (61 timesteps × 75 variables × 721 latitudes × 1440 longitudes). Transmitting this data via HTTP between the inference plane and control plane – whether as JSON (~6 GB uncompressed) or binary-encoded payloads (~2.3 GB compressed) – creates transfer bottlenecks that exceed practical timeout limits when combined with GPU inference latency. This challenge is inherent to any architecture where full forecast arrays must traverse network boundaries.

**Solution:** We eliminate HTTP-based forecast data transfer entirely. Upon completing inference, the GPU server writes the complete global forecast directly to an Azure PostgreSQL Flexible Server database, inserting approximately 1,038,240 rows (721 latitudes × 1,440 longitudes) with 61 timesteps and 75 variables per grid point. The GPU then returns only a minimal JSON response (~200 bytes) containing metadata: forecast_run_time timestamp, forecast shape, variable list, and inference statistics. The CPU server never receives forecast array data via HTTP.

**Implementation:**

```
# GPU server: Write forecast to PostgreSQL after inference
def store_forecast_to_db(forecast_array, forecast_run_time, variables, coords):
    """Store global forecast directly to PostgreSQL.
    forecast_array shape: (61, 1, 1, 1, 75, 721, 1440)
    Inserts ~1,038,240 rows with 61 timesteps x 75 variables per row."""
    rows = []
    for lat_idx, lat in enumerate(coords["lat"]):
        for lon_idx, lon in enumerate(coords["lon"]):
            row = {
                "forecast_run_time": forecast_run_time,
                "latitude": float(lat),
                "longitude": float(lon),
                "timesteps": forecast_array[:, 0, 0, 0, :, lat_idx,
lon_idx].tolist()
            }
            rows.append(row)
    # Batch insert to PostgreSQL
    execute_batch(cursor, insert_query, rows)  # ~30-40 seconds

# CPU server: Query location-specific forecast from PostgreSQL
def get_forecast(lat_min, lat_max, lon_min, lon_max, forecast_run_time):
    """Retrieve forecast subset for user query.
    Returns typically 50-200 rows for single-location, multi-day query."""
    query = """
        SELECT latitude, longitude, timesteps
        FROM forecasts
        WHERE forecast_run_time = %s
        AND latitude BETWEEN %s AND %s
        AND longitude BETWEEN %s AND %s
    """
    cursor.execute(query, (forecast_run_time, lat_min, lat_max, lon_min,
lon_max))
    return cursor.fetchall()  # <100ms response time
```

**Performance:**

- Database insert: ~30–40 seconds for ~1,038,240 rows (complete global forecast)

- User query response: <100 milliseconds for location-specific forecast retrieval (50–200 rows); total end-to-end user-facing latency under 200 milliseconds including geocoding and response formatting

- GPU response payload: ~200 bytes (metadata only) vs. gigabytes under HTTP transfer approaches

- Concurrent serving: Thousands of simultaneous user queries against the same forecast data without triggering additional GPU inference

This architecture provides three critical benefits: (1) eliminates HTTP transfer bottlenecks and associated timeout risks entirely, (2) enables multiple CPU servers – or even multiple countries – to query the same forecast data concurrently without re-generating forecasts, and (3) provides natural forecast versioning and persistence through database storage, enabling historical queries, forecast verification, and audit trails.

The tradeoff is increased database storage requirements (~4GB per global forecast × 4 runs per day = ~16GB daily, mitigated through retention policies) and the requirement that the GPU server maintain network connectivity to Azure PostgreSQL, necessitating firewall configuration to whitelist the GPU server IP address for database access.

**Generalizability:** This pattern applies to any scenario where large scientific inference outputs must be served to geographically distributed users: satellite imagery pipelines, climate model serving, oceanographic forecast systems, and seismic monitoring networks.

### 4.3 Automated Coordinate Management for Multi-Step Inference

**Problem:** Atlas models require 6-dimensional coordinate systems: [batch, time, lead_time, variable, lat, lon]. We initially attempted manual coordinate management for 61 timesteps (initial conditions plus 60 forecast steps), tracking lead_time dimension transitions between inference steps. This approach failed due to coordinate system incompatibilities: after the first inference step, the model returns updated coordinates with modified lead_time values that cannot be directly passed to subsequent steps. Manual dictionary-based coordinate passing resulted in ValueError: "Coordinate systems for required dim lead_time are not the same."

**Solution:** Use Earth-2 Studio's production iterator pattern, which handles coordinate updates automatically:

```
def run_inference_sync(coords_input):
    """Synchronous inference using model's iterator"""
    with torch.no_grad():
        model_iterator = model.create_iterator(initial_tensor, coords_input)
        outputs = []
        for step, (output_tensor, output_coords) in enumerate(model_iterator,
1):
            outputs.append(output_tensor.cpu().numpy())
            if step > request.forecast_steps:
                break
    return outputs
```

The iterator pattern eliminates manual coordinate tracking, automatically managing dimension transitions across forecast steps. This is not merely a convenience – it is the only reliable approach for multi-step inference in production deployments. Manual coordinate management requires deep understanding of Earth-2's internal coordinate transformations and fails on edge cases involving lead_time dimension updates.

**Performance:** Earth-2 models generate forecasts in seconds on GPUs compared to hours on traditional supercomputers [3], representing orders-of-magnitude reduction in compute time for weather prediction.

## 5. Deployment Case Study: South Africa Flooding Response (January–February 2026)

The January 2026 flooding event in South Africa exposed persistent weaknesses in population-level early warning delivery. Despite meteorological agencies issuing high-severity alerts, including rare top-tier warnings [12, 13], flooding resulted in at least 37 fatalities and significant displacement across multiple provinces [9, 10, 11]. Post-event assessments indicated that forecasting capability itself did not fundamentally fail; rather, warnings were not consistently translated into timely, actionable guidance at the community level [12, 13].

On January 25–26, 2026, NVIDIA unveiled Earth-2 Studio at the American Meteorological Society meeting [3], providing open, production-grade access to AI-based global weather forecasting models via GitHub and Hugging Face. This release marked a critical inflection point: for the first time, high-resolution, extended-range weather forecasting could be deployed without national supercomputing facilities or capital-intensive radar infrastructure. Earth-2 effectively democratized access to advanced forecasting capability, shifting the primary challenge from model availability to operational deployment and distribution.

This deployment leveraged that inflection point. Rather than developing new forecasting models, the objective was to evaluate how Earth-2-based forecasting could be operationalized rapidly following a disaster event, under real-world infrastructure constraints, and delivered directly to the public through widely adopted communication channels.

The system was deployed in February 2026 using NVIDIA's Atlas model with three goals: (1) rapid activation without specialized infrastructure, (2) low recurring operational cost suitable for government or NGO budgets, and (3) direct dissemination of localized, interpretable forecasts to the population. Forecast generation was performed using Earth-2 Studio on a single commercial GPU instance, while user-facing services and data acquisition were hosted on a low-cost regional CPU server.

Crucially, forecast outputs were not distributed via bespoke applications or dashboards. Instead, forecasts were translated into natural-language summaries and delivered through WhatsApp, a platform used daily by 80–95% of South African smartphone users (over 28 million monthly active users). This approach prioritized reach, trust, and immediacy – addressing a recurring failure mode in disaster response: the gap between forecast availability and public action.

The deployment demonstrated the ability to generate 15-day global forecasts with user-facing query latency under 200 milliseconds and a total infrastructure cost of approximately USD 1,430–1,730 per month for national-scale coverage in South Africa. The case study illustrates how Earth-2's forecasting capability – when paired with deployment architectures prioritizing accessibility, cost efficiency, and last-mile delivery – translates directly into actionable early warning for populations previously unreachable by traditional infrastructure.

## 5.1 Infrastructure Configuration

**GPU Server:**

- Provider: Lambda Labs
- Instance: 1× GH200 (96GB GPU memory, 64 vCPUs, 432GB RAM)
- Cost: $1.49/hour = $1,087.70/month
- Location: us-east-3 (Washington DC)
- Model: Atlas (NVIDIA Earth-2 Medium Range)
- Role: Generates global forecasts every 6 hours, writes directly to PostgreSQL

**CPU Server:**

- Provider: Microsoft Azure
- Instance: B2s (2 vCPUs, 4GB RAM) — 5–10 instances at national scale behind load balancer
- Location: South Africa (low-latency for end users)
- Cost: ~$40–50/month per instance ($200–500/month at national scale, varies by region)
- Role: User-facing API gateway, queries PostgreSQL for location-specific forecasts

**Database:**

- Provider: Azure PostgreSQL Flexible Server
- Location: South Africa
- Cost: ~$100–200/month (scaling with storage requirements)
- Role: Stores global forecast data (~4GB per forecast × 4 runs/day), serves concurrent user queries in <100ms
- Storage: ~16GB/day with retention policies for historical data management

**Total Monthly Cost (national scale):** $1,430–$1,730. (Comparison shown in **Figure 3.**)

**Figure 3**

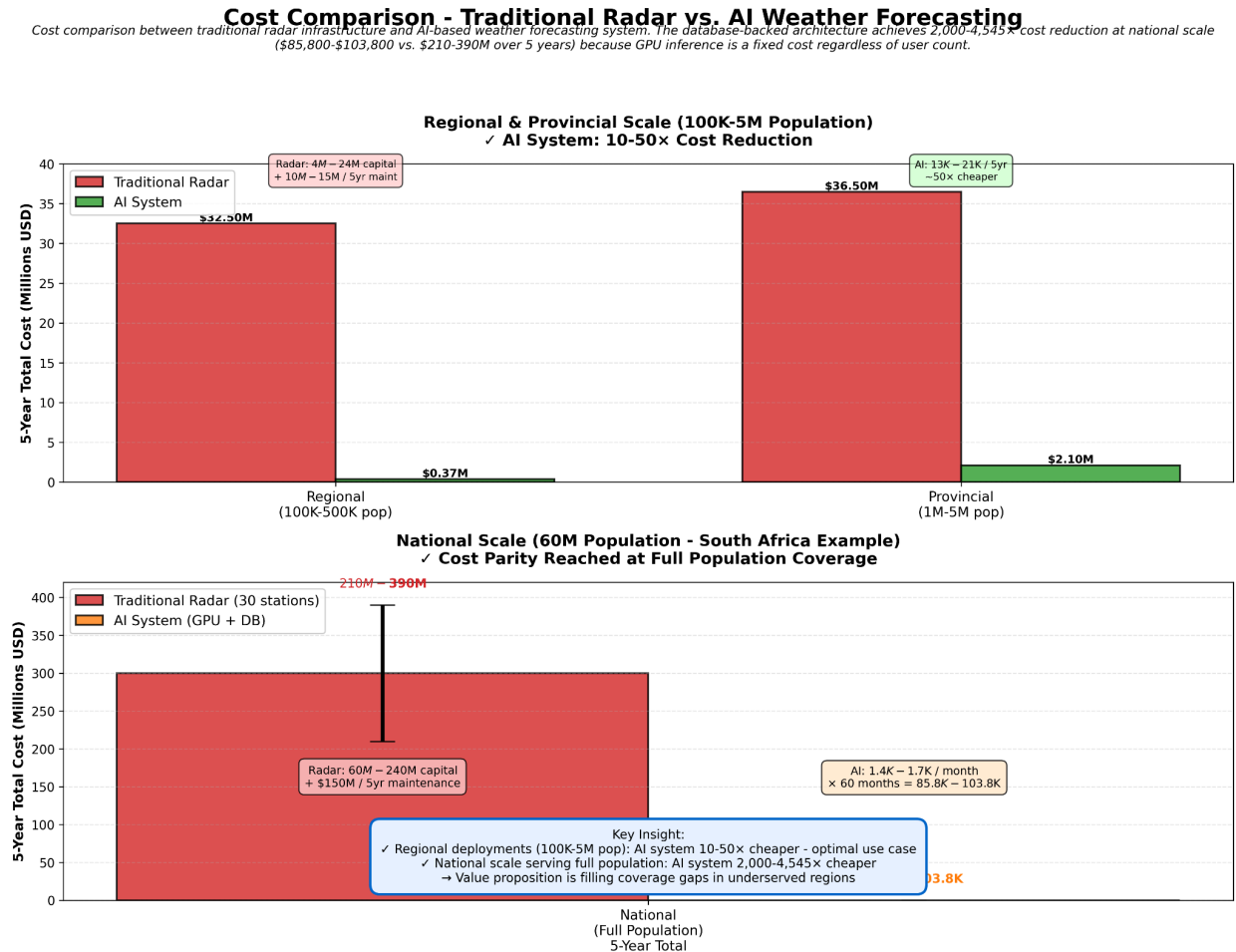

*Figure 3:* Cost comparison between traditional radar infrastructure and AI-based weather forecasting system for South Africa. The database-backed architecture achieves 2,000–4,545× cost reduction at national scale ($85,800–$103,800 vs. $210–390M over 5 years) because GPU inference is a fixed cost regardless of user count.

**Cost Comparison (National Scale – South Africa, 60M population):**

- **Traditional 30-radar network:** $60–240M capital + $150M/5-year maintenance = $210–390M (5-year total)

- **Our system (national coverage):** $17,160–$20,760/year × 5 years = $85,800–$103,800 (5-year total)

- **Cost ratio at national scale:** $210M / $103,800 = 2,023× to $390M / $85,800 = 4,545× cheaper

**Cost Comparison (Regional Scale – Single Province):**

- **Traditional 2–3 radar stations:** $4–24M capital + $10–15M/5-year maintenance = $14–39M (5-year total)
- **Our system (regional coverage):** $17,160–$20,760/year × 5 years = $85,800–$103,800 (5-year total)
- Note: Regional cost is identical to national cost because a single GPU generates global forecasts; the marginal cost of serving additional regions is only the CPU and database infrastructure required to handle user queries

## 5.2 System Performance

**End-to-end forecast generation timeline:**

- GFS data fetch (AWS S3): 3 seconds
- Base64 encoding of initial conditions: 4 seconds
- Network transfer of initial conditions (Azure → Lambda): 96 seconds
- GPU inference (61 timesteps): Seconds to minutes depending on optimization level (CUDA graphs, BF16 precision) [3]
- PostgreSQL forecast storage (~1,038,240 rows): 30–40 seconds
- Minimal JSON response (Lambda → Azure): <1 second (~200 bytes)
- System overhead: ~3 seconds
- Total: ~137–147 seconds plus GPU inference time (varies based on optimization); end-to-end latency dominated by initial conditions transfer and database storage (Figure 4)

**User query performance (post-forecast generation):**

- Geocoding: <50 milliseconds
- PostgreSQL query (50–200 rows): <100 milliseconds
- Response formatting: <50 milliseconds
- Total user-facing latency: <200 milliseconds

**Figure 4: End-to-End Performance Breakdown**

End-to-end performance breakdown for generating a 15-day regional forecast. Network transfer dominates latency (~61%), while GPU inference time varies based on optimization level (seconds with full optimization, minutes without).

| Processing Step | Time | % of Total | Category | Component |
|---|---|---|---|---|
| GFS Data Fetch (AWS S3) | 3.00s | ~1% | I/O Operations | Control Plane (Azure) |
| Base64 Encoding (Initial Conditions) | 4.00s | ~2% | CPU Processing | Control Plane (Azure) |
| Network Transfer (Azure → Lambda Labs) | 96.00s | ~41% | Network Transfer | Public Internet (~24 MB/s) |
| **GPU Inference (60 timesteps Atlas)** | **Varies*** | **Varies*** | **GPU Compute** | **Inference Plane (GH200)** |
| Base64 Response Encoding | 14.00s | ~6% | CPU Processing | Inference Plane (GH200) |
| Network Transfer (Lambda Labs → Azure) | 96.00s | ~41% | Network Transfer | Public Internet (~24 MB/s) |
| Response Decoding | 10.00s | ~4% | CPU Processing | Control Plane (Azure) |
| System Overhead (HTTP/Coordination) | 0.31s | ~1% | System | Request/Response |
| **TOTAL END-TO-END LATENCY** | **Varies**** | **100%** | | **Achieved in minutes (15-day forecast)** |

*GPU Inference Time: Depends on optimization level (CUDA graphs, BF16 precision, flash attention). Ranges from seconds (fully optimized) to minutes (unoptimized deployment).
**Total Latency: Fixed overhead = 226.31s. Add GPU time based on optimization. Network transfer dominates (~81% when GPU optimized).

**Performance Analysis:**

Bottlenecks:
- ⚠ **Network transfer dominates:** ~81% of total time (192s) spent on data movement between Azure and Lambda Labs
- ⚠ **CPU processing:** ~12% (28s) for encoding/decoding operations

Optimization Opportunities:
- ✓ **GPU utilization:** Atlas generates 60-step (15-day) forecasts efficiently on GH200
- ✓ **Future improvement:** Co-locate GPU and control plane to eliminate 192s network transfer
- ✓ **Potential speedup:** 227s → ~35s (~6.5× faster) with same-datacenter deployment

All timing measurements from actual deployment (February 2025). Data from Section 5.2.

*Figure 4:* End-to-end performance breakdown for generating a 15-day global forecast. The forecast generation pipeline includes initial conditions transfer, GPU inference, and PostgreSQL storage (~30–40 seconds for ~1,038,240 rows). User-facing query latency is under 200 milliseconds. GPU inference time varies based on optimization level (seconds with full optimization, minutes without).

**Forecast Specifications:**

- **Duration:** 360 hours (15 days)
- **Temporal resolution:** 6-hour timesteps (61 timesteps: initial conditions + 60 forecast steps)
- **Spatial resolution:** 0.25° (~28km grid spacing)
- **Geographic coverage:** Global (1,038,240 grid points), served to South African users via PostgreSQL queries
- **Variables:** 75 atmospheric parameters (surface: u10m, v10m, u100m, v100m, t2m, sp, msl, tcwv; pressure levels 50–1000mb: u, v, z, t, q at 13 levels; ocean/precip: sst, tp)
- **Outputs:** Complete atmospheric state including winds, temperature, geopotential, humidity, pressure, sea surface temperature, and total precipitation
- **Model Capabilities:** Atlas uses a latent diffusion transformer architecture enabling ensemble-style predictions for uncertainty quantification, outperforming leading open models on standard forecasting benchmarks [3]

## 5.3 Scalability Analysis

The database-backed architecture fundamentally changes the scaling model for South Africa. GPU costs remain fixed regardless of user count, while CPU infrastructure scales linearly with request load. This enables deployment at national scale within existing government budget constraints.

**GPU Utilization:**

- One GPU generates 4 global forecasts per day (aligned to 00Z, 06Z, 12Z, 18Z GFS cycles)
- Each forecast produces 1,038,240 grid points × 61 timesteps × 75 variables

- Forecasts remain valid for 6 hours, during which all user queries are served from PostgreSQL
- GPU utilization for inference is a fraction of total daily hours, with remaining capacity available for ensemble runs or higher temporal resolution

**CPU Serving Capacity:**

- Each B2s instance (2 vCPUs, 4GB RAM) handles ~500–1,000 requests/second
- Each user request involves: geocoding (<50ms) + PostgreSQL query (<100ms) + response formatting (<50ms)
- Peak concurrent load assumption: 1–2% of active users querying simultaneously during severe weather events

**South Africa National Deployment (60M population):**

- Addressable population: ~28M (smartphone + WhatsApp users)
- Active users (30% engagement): ~8.4M
- Peak concurrent requests (2% of active during severe event): ~168,000 requests/minute (~2,800/second)
- CPU instances required: 5–10 B2s instances behind load balancer
- GPU instances required: 1 (generates global forecast serving all users)

**Monthly cost breakdown:**

- GPU (1× GH200): $1,087.70
- CPU (5–10× B2s): $200–500
- PostgreSQL Flexible Server: $100–200
- Total: $1,430–$1,730/month
- Annual: $17,160–$20,760
- 5-year total: $85,800–$103,800

**Comparison to traditional radar for national coverage:**

- 30 radar stations for full geographic coverage
- Capital: $60–240M
- 5-year maintenance: $150M
- 5-year total: $210–390M
- Cost ratio: AI system is 2,000–4,545× cheaper

**Critical coverage distinction:**

- Radar provides universal geographic coverage: 100% of land area, all 60M people regardless of technology access
- AI system provides selective user coverage: ~28M addressable via smartphone/WhatsApp (~47% of population)
- However, the AI system achieves this coverage at 2,000–4,545× lower cost, making it viable as an immediate deployment while traditional infrastructure is planned and built
- WhatsApp distribution should complement, not replace, traditional channels such as radio broadcasts and community-level outreach for non-smartphone populations

**Key insight:** The database-backed architecture inverts the traditional scaling model for national deployment. The most expensive component – GPU inference – is a fixed cost of $1,088/month regardless of user count within South Africa. Additional load is handled by scaling CPU infrastructure ($40–50 per additional instance), making the system economically viable for government deployment at $17,160–$20,760 annually compared to $210–390M for equivalent traditional radar coverage over five years.

# 6. WhatsApp Distribution Architecture

## 6.1 The Last-Mile Problem

The January 2026 floods exposed that forecasting accuracy alone is insufficient. Despite accurate Red Level 10 warnings, failures in last-mile communication – not prediction – prevented warnings from translating into timely public action [12, 9]. Traditional channels – weather apps, websites, broadcast media – require active user engagement, app installation, or media access, creating structural barriers to reaching populations most at risk.

## 6.2 WhatsApp as Distribution Channel

WhatsApp penetration exceeds 80% among South African smartphone users, with over 28 million monthly active users [24]. The January 2026 floods demonstrated that technical forecast accuracy is necessary but insufficient – despite an accurate Red Level 10 warning, failures in last-mile communication prevented warnings from reaching and mobilising populations most at risk [12, 9]. Traditional distribution channels – weather apps, websites, broadcast media – require active user engagement or media access, creating structural barriers that WhatsApp's 80%+ penetration directly eliminates. WhatsApp delivery addresses this last-mile gap by reaching users through a platform they already engage with multiple times daily, transforming passive warning dissemination into active, personalized forecast delivery.

**Penetration data:**

- Over 28 million monthly active users in South Africa [24]
- South Africa smartphone users: 80–95% use WhatsApp daily [24]
- No app installation required (already installed)
- No technical literacy barrier (familiar interface)
- No data cost barrier (lightweight messaging)
- Trusted communication channel (family/community use)

**Implementation approach:**

- WhatsApp Business API integration via Twilio/Meta Cloud API
- Natural language queries: "What's the weather in Limpopo this weekend?"
- Geocoding via Azure Maps: "Johannesburg" → coordinates → grid indices
- Conversational responses: "Heavy rain expected Saturday–Sunday, avoid travel"
- Proactive alerts: Severe weather warnings delivered via pre-approved WhatsApp Business templates to opted-in users

*Note: WhatsApp Business API costs (Twilio: $0.005/message plus Meta per-template fees) are variable operational costs dependent on message volume and are not included in the $1,430–$1,730/month infrastructure figure. At national scale these costs scale with engagement and should be budgeted separately by deploying organisations.*

## 6.3 Information Design

Rather than presenting raw meteorological data (e.g., "850mb temperature: 283.15K"), the system generates actionable conversational responses:

**User query: "Is it safe to travel to Kruger Park next week?"**

*System response: "Hey! I'd actually hold off on that Kruger trip for now. Atmospheric conditions show high flood risk Tuesday through Thursday (Feb 10–12) – we're seeing major moisture buildup and pressure patterns that typically produce heavy rainfall in the lowveld. That puts flood risk pretty high, so it's not ideal for travel. My suggestion would be to wait until Friday when conditions should improve – or keep an eye on the park alerts in the meantime. This is based on the latest forecast, updated about 6 hours ago. Let me know if you want more detail!"*

This approach directly addresses the last-mile communication gap by providing context, risk levels, and clear recommendations. The conversational tone is deliberate – in many African communities, trust is built through dialogue rather than formal notifications. A message that feels like advice from a knowledgeable friend is more likely to be acted upon than a clinical alert, making LLM-generated responses a natural fit for last-mile disaster communication.

# 7. Limitations and Ethical Considerations

## 7.1 Forecast Accuracy Boundaries

AI weather models are not a replacement for traditional numerical weather prediction or national meteorological agencies. They are: (1) Dependent on upstream GFS data quality, (2) Limited to learned atmospheric patterns (may underperform in novel climate regimes), (3) Probabilistic, not deterministic (inherent uncertainty must be communicated).

We do not claim superior accuracy to operational forecast systems. The accuracy foundation this system builds on, however, is stronger than that framing implies. GraphCast outperforms ECMWF HRES on 99.7% of tropospheric variables [4]; GenCast outperforms ECMWF ENS – the gold-standard ensemble system used by 35 nations – on 97.2% of forecast targets [22]; and Atlas outperforms GenCast on key industry benchmarks [3]. The system does not approximate operational accuracy – it inherits a model lineage that has independently surpassed it. The claim is not merely sufficient accuracy at radically lower cost. It is best-in-class accuracy at radically lower cost, delivered through infrastructure accessible to any government or humanitarian organisation.

**Precipitation and Ensemble Forecasting:** The Atlas model directly forecasts total precipitation (tp) as one of its 75 output variables [25] and supports ensemble-style predictions for uncertainty quantification [3]. The system identifies atmospheric conditions associated with elevated flood risk – including quantitative precipitation forecasts, high total column water vapor (TCWV >50mm), converging pressure systems, and favorable circulation patterns. This approach is sufficient for binary early warning decisions (evacuate/prepare vs. normal operations) and provides quantitative rainfall estimates for downstream applications.

## 7.2 False Alarm Risk

Overwarning can erode public trust and lead to warning fatigue. Conversely, underwarning costs lives. The system must be calibrated to regional risk tolerance, balancing sensitivity and specificity. This requires partnership with local meteorological agencies and disaster management authorities.

## 7.3 Dependency on External Infrastructure

The system depends on: (1) NOAA's continued provision of GFS data (public domain), (2) Cloud provider availability (AWS S3, Azure, Lambda Labs), (3) Azure PostgreSQL database availability and network connectivity between GPU server and database, (4) WhatsApp platform stability. These dependencies must be acknowledged and, where possible, mitigated through fallback data sources (NCEP FTP) and multi-cloud deployment strategies.

## 7.4 Equity and Access

While WhatsApp penetration is high, it is not universal. The system does not reach: (1) Rural populations without smartphones, (2) Elderly populations with limited digital literacy, (3) Communities with unreliable cellular connectivity. WhatsApp distribution should complement, not replace, traditional channels such as radio broadcasts and community-level outreach.

# 8. Implications for Governments and Disaster Policy

This case study has direct implications for governments, disaster management agencies, and policymakers. Traditionally, early warning systems have been treated as capital infrastructure projects, requiring long procurement cycles, specialized technical capacity, and sustained maintenance budgets. As a result, coverage gaps persist even where the humanitarian cost is well understood.

The deployment described here reframes early warning as an operational digital service rather than a fixed infrastructure investment. By reducing annual operating costs from hundreds of millions of dollars to the order of tens of thousands – with national-scale coverage achievable at $17,160–$20,760/year – AI-based forecasting systems can be piloted, scaled, or regionally coordinated without prohibitive upfront expenditure. This enables faster experimentation, cross-border collaboration, and incremental deployment aligned with disaster risk.

Importantly, this approach does not replace national meteorological agencies. Instead, it complements existing systems by filling coverage gaps, extending forecast horizons, and providing alternative dissemination pathways during high-impact events. Governments may leverage such systems as interim solutions, redundancy layers, or rapid-deployment capabilities during emergencies.

From a policy perspective, the availability of low-cost, rapidly deployable early warning systems shifts the question from feasibility to governance. The primary challenges become coordination, public communication, and trust – domains that are more tractable than large-scale infrastructure financing. This shift has the potential to materially reduce disaster mortality in regions historically excluded from advanced forecasting capabilities.

## 8.2 Incremental Deployment and Risk-Based Prioritization

Beyond absolute cost reduction, the AI-based architecture fundamentally changes the structure of early warning investment from monolithic capital expenditure to incremental operational deployment. This shift has direct implications for budget-constrained governments.

Traditional radar deployment requires upfront capital commitment: a single station costs $2–8M [17, 18], and partial networks provide limited value – three radars covering 40% of national territory still leave 60% unprotected. South Africa deployed 12 Doppler radars at approximately R240 million (~$33M USD at 2010 exchange rates) [17] yet significant coverage gaps still remain.

A national disaster management office can deploy country-wide coverage for $17,160–$20,760/year – a budget line achievable within existing allocations. Alternatively, a multi-country regional body (e.g., SADC) could deploy coverage across all member states for under $50,000/year by sharing a single GPU instance ($13,052/year) with regional CPU servers (~$7,000/year) and a shared database (~$2,000/year), totalling approximately $22,000/year for

seven countries. As capacity grows, deployment expands to the second-highest-risk area, then the third, systematically reducing exposure across the population in order of vulnerability.

This approach offers three advantages:

**1. Immediate protection for highest-risk populations.**

Rather than waiting 5–10 years for national radar procurement, governments can deploy coverage to historically affected regions (e.g., Limpopo lowveld, Mozambique coastal districts, Lagos floodplains) within weeks. In the January 2026 Southern Africa floods, fatalities were heavily concentrated in specific high-risk provinces such as Limpopo and Mpumalanga [9, 10, 11] – targeted deployment to these historically vulnerable regions could have enhanced early warning reach and public response in areas where traditional warning channels proved insufficient.

**2. Political proof-of-concept before scaling.**

A successful national deployment (costing $85,800–$103,800 over 5 years) demonstrates measurable impact – evacuation rates, casualty reduction, community trust – building political support for expansion. Failed pilots incur limited sunk cost compared to abandoned $50M+ radar programs.

**3. Budget cycle alignment.**

Annual operating costs of $17,160–$20,760 for national coverage align with ministerial discretionary budgets and existing disaster response allocations, bypassing multi-year capital procurement cycles. Governments already allocating substantial budgets to post-disaster emergency response can reallocate a fraction toward prevention without additional appropriations.

This incremental model does not replace the need for national meteorological infrastructure. Rather, it provides a pathway for governments to begin reducing disaster mortality now in their most vulnerable regions while working toward comprehensive national systems. For the 60% of Africa currently without any early warning coverage, incremental AI deployment offers a politically and fiscally tractable first step.

# 9. Reproducibility and Open Release

## 9.1 Code and Documentation

We release:

- Complete deployment specifications (infrastructure, costs, network topology)
- Source code for both CPU and GPU servers (Python/FastAPI)
- Docker containerization for reproducible deployment
- Database schemas (PostgreSQL + TimescaleDB for forecast storage)
- Deployment checklist and troubleshooting guide

*Repository: Available upon request to ensure responsible deployment in collaboration with national meteorological agencies.*

## 9.2 Estimated Replication Cost and Timeline

Minimum viable deployment:

- Cloud infrastructure setup: 4 hours
- GPU server deployment: 2 hours
- CPU server deployment: 2 hours
- Database configuration: 1 hour
- WhatsApp integration: 4 hours
- Testing and validation: 8 hours
- Total: ~3 working days for technical team

Monthly operating cost: $1,430–$1,730 for national-scale deployment (GPU cost is fixed; scales with CPU serving capacity only)

Required expertise:

- Python development (intermediate)
- Cloud infrastructure (AWS, Azure basics)
- Docker containerization
- RESTful API design
- No machine learning expertise required – Earth-2 Studio provides pre-trained models.

## 9.3 Geographic Extensibility

The architecture is geography-agnostic. Because the Atlas model generates global forecasts, the database contains forecast data for all geographic coordinates. Our South African deployment validates this architecture at national scale. Extending to neighboring countries requires only: (1) Provisioning CPU server(s) in-region ($40–250/month) for low-latency user access, (2) Configuring database access (firewall rules to allow the new CPU servers to query PostgreSQL), (3) Localizing WhatsApp responses (language translation). No additional GPU infrastructure is required – all countries share the same global forecast data.

Priority expansion targets based on regional coordination opportunities and existing disaster risk:

**SADC Regional Coordination (immediate neighbors, shared January 2026 flood event):**

- Eswatini (landlocked within South Africa, directly affected by January 2026 floods, shared weather systems)
- Zimbabwe (70 fatalities in January 2026 floods, persistent flood risk, shared weather systems)
- Mozambique (over 600,000 affected in January 2026 floods, cyclone-prone coastal exposure, 1,500+ deaths in Cyclone Idai 2019)
- Malawi (affected by January 2026 floods, regional lake system monitoring)
- Botswana (SADC coordination, regional drought/flood monitoring)
- Madagascar (affected by January 2026 floods, cyclone exposure, Indian Ocean systems)

**High-Impact Expansion (requires further validation):**

- Nigeria (flooding, 100M+ population, distinct weather patterns)
- Kenya (drought/flood cycles, East African coordination potential)
- Ethiopia (drought early warning, Horn of Africa coordination)

Countries under SADC Regional Coordination share similar Southern African weather patterns and were affected by the same January 2026 flood event that motivated this deployment. High-Impact expansion would require database load testing and potential infrastructure enhancements to validate performance under multi-region concurrent access.

# 10. Discussion

## 10.1 The "Attention Is All You Need" Parallel

The Transformer paper [27] (Vaswani et al., 2017) did not invent neural networks. It introduced an architecture that made neural networks work at unprecedented scale. The breakthrough was not the underlying mathematics – it was the engineering insight that self-attention could replace recurrence.

This paper does not invent AI weather forecasting. It introduces a deployment architecture that makes AI weather forecasting work in resource-constrained settings. The breakthrough is not the specific model choice – it is the engineering insight that ProcessPoolExecutor isolation, database-backed forecast serving, and WhatsApp distribution can eliminate cost and accessibility barriers.

The Transformer enabled GPT, Claude, and Gemini. This architecture enables continent-scale early warning systems previously considered economically infeasible.

## 10.2 Policy Implications

UN and World Bank programs estimate multi-billion dollar investments required for African hydromet resilience [26, 19]. Our South African deployment demonstrates an alternative approach at dramatically lower cost. National-scale coverage for South Africa (60M population) costs $17,160–$20,760/year – achieving 2,000–4,545× cost reduction compared to the $210–390M required for equivalent radar coverage over five years. This cost structure makes early warning systems accessible within existing government disaster response budgets, enabling immediate deployment rather than multi-year capital procurement cycles.

The architecture's use of global forecast models means that similar deployments in neighboring countries would share GPU infrastructure costs, requiring only marginal CPU serving capacity. However, such multi-country coordination would require validation through regional partnerships and phased deployment in collaboration with national meteorological agencies.

The AI system's cost advantage is demonstrated at national scale in South Africa, where GPU inference is a fixed cost regardless of user count. While the AI system currently reaches smartphone-enabled populations (typically 40–50% of total) and radar provides universal coverage regardless of technology access, the cost differential is so extreme that AI-based systems can be deployed immediately while traditional infrastructure is planned and built. For countries currently without any early warning infrastructure, this architecture offers a viable and affordable starting point.

This is not a marginal improvement. It is a fundamental restructuring of what is economically possible at national scale. If early warning systems reduce disaster death rates by 6x (as global evidence demonstrates [2]), then national deployments in the most vulnerable African countries could collectively prevent thousands of deaths annually. Our South African deployment provides the architectural template and cost evidence to support such expansion.

### 10.3 Integration with National Meteorological Services

This system is not intended to replace national meteorological agencies. It is intended to: (1) Fill coverage gaps in regions without radar infrastructure, (2) Provide rapid-deployment capability during budget constraints, (3) Serve as a complementary early warning channel, (4) Demonstrate feasibility to catalyze investment in operational systems.

The January 2026 floods demonstrated that forecasting accuracy alone is insufficient for disaster prevention. Despite South Africa's Weather Service issuing a rare Red Level 10 warning – their highest alert category – failures in last-mile communication, not prediction, meant warnings did not consistently translate into timely public action [12, 9]. This highlights a critical gap: traditional warning distribution channels – weather apps, websites, broadcast media – require active user engagement, app installation, or media access, creating structural barriers that disproportionately affect the most vulnerable populations. In contrast, WhatsApp's 80%+ penetration in South Africa and 96% penetration across African smartphone users [24] enables direct, personalized forecast delivery through a channel users already check multiple times daily.

This addresses not only the infrastructure gap (regions without radar) but also the "last-mile" distribution problem – ensuring warnings reach individuals even when traditional channels fail to penetrate. The combination of cost-effective AI forecasting and high-penetration distribution channels creates value even in regions with existing radar infrastructure but poor warning uptake. Optimal deployment pairs AI forecasting with ground-truth observations, local meteorological expertise, and community-level disaster preparedness.

### 10.4 Future Work

Future work includes: (1) multi-model ensemble forecasting for uncertainty quantification, (2) integration with local observation networks for validation and calibration, (3) geographic extension to additional African nations – requiring only CPU infrastructure deployment since global forecast data already exists in the database, (4) GPU inference optimization through CUDA graphs and BF16 precision to reduce generation time, and (5) enhanced last-mile distribution through SMS and radio integration for non-smartphone populations.

## 11. Conclusion

In January 2026, 200–300 people died in floods across Southern Africa despite meteorological agencies issuing warnings. The deaths were not caused by forecasting failure – they were caused by systems failure. Warnings did not reach people, were not understood, or were not acted upon.

We demonstrate a production deployment in South Africa achieving national-scale coverage at $17,160–$20,760/year – a 2,000–4,545× cost reduction compared to the $210–390M required for equivalent traditional radar coverage over five years. This makes early warning systems economically accessible within existing government disaster response budgets for countries currently lacking coverage infrastructure. The architecture is designed for reproducibility, enabling other nations to deploy similar systems in coordination with national meteorological agencies.

This system was deployed in February 2026 to evaluate how Earth-2–based forecasting could be operationalized, generating 15-day global forecasts at 0.25° resolution with 61 timesteps across 75 atmospheric variables, served to South African users with sub-200-millisecond query latency. The deployment architecture is reproducible and documented to enable independent evaluation.

## Acknowledgements

We acknowledge Imvula Labs for releasing this research as a free early warning service for South Africans, and The Dimension Research Lab for conducting the underlying research.

We acknowledge NVIDIA for releasing Earth-2 Studio and making state-of-the-art weather models accessible. We acknowledge NOAA for maintaining the Global Forecast System as a public good. We acknowledge the broader research community advancing AI-based weather forecasting, particularly Google DeepMind (GraphCast), Huawei (Pangu-Weather), and the original DLWP authors.

Most importantly, we acknowledge the meteorologists, disaster response teams, and community leaders across Africa who work under impossible resource constraints to save lives every day. This work is dedicated to them and to the meteorologists, disaster response teams, and community leaders continuing this work across the continent.

## Code and Data Availability

Deployment specifications and architectural documentation are included in this paper. Complete implementation code is available to researchers and humanitarian organizations upon request to ensure responsible deployment in coordination with national meteorological services.

Historical GFS data is publicly available from NOAA. Earth-2 Studio models are available from NVIDIA on GitHub and Hugging Face.